\documentclass[aps,prl,twocolumn,citeautoscript,showpacs]{revtex4-1}
\usepackage{graphicx}
\usepackage{amssymb}
\usepackage{amsmath}
\usepackage{mathrsfs}
\usepackage{amsfonts}

\newcommand{\bk}{\textbf{k}}
\newcommand{\bx}{\textbf{x}}

\newcommand{\bp}{\textbf{p}}
\newcommand{\bq}{\textbf{q}}

\newcommand{\bQ}{\textbf{Q}}
\newcommand{\bP}{\textbf{P}}
\newcommand{\bK}{\textbf{K}}

\newcommand{\bhk}{\hat{\textbf{k}}}
\newcommand{\bhn}{\hat{\textbf{n}}}

\newcommand{\ud}{\mathrm{d}}
\newcommand{\vare}{\varepsilon}
\renewcommand{\Re}{\textrm{Re}}

\newcommand{\vect}[1]{\textbf{#1}}
\newcommand{\Pdd}[3][]{\frac{\partial^{#1} #3}{\partial #2^{#1}}}

\newcommand{\pdd}[3][]{\partial^{#1} #3 / \partial #2^{#1}}

\DeclareMathOperator{\sgn}{sgn}

\begin{document}

\title{Shear Viscosity in a Non-Fermi Liquid Phase of a Quadratic Semimetal}
\author{Philipp T.~Dumitrescu}
\affiliation{Department of Physics, University of California, Berkeley, CA 94720, USA}
\email{philippd@berkeley.edu}
\date{\today}

\begin{abstract}
We study finite temperature transport in the Luttinger-Abrikosov-Beneslavskii phase -- an interacting, scale invariant, non-Fermi liquid phase found in quadratic semimetals. We develop a kinetic equation formalism to describe the d.c.~transport properties, which are dominated by collisions, and compute the shear viscosity $\eta$. The ratio of shear viscosity to entropy density $\eta/s$ is a measure of the strength of interaction between the excitations of a quantum fluid. As a consequence of the quantum critical nature of the system, $\eta / s$ is a universal number and we find it to be consistent with a bound proposed from gauge-gravity duality.
\end{abstract}


\maketitle

Transport coefficients such as the electrical and thermal conductivities or the viscosities play a central role in describing condensed matter systems. They are experimentally measurable and contain signatures which characterize the different phases of matter. Nonetheless, for strongly interacting systems at finite temperatures, as in the vicinity of a quantum critical point, there remain difficulties in calculating transport coefficients analytically due to the lack of a quasiparticle description \cite{Sachdev:2011kq}. Recently, insight has been gained from the relationship between strongly coupled field theories and classical gravitational theories in the context of gauge-gravity duality \cite{Sachdev:2012sf, Hartnoll:2009xy}. In particular, Kovtun, Son and Starinets \cite{Kovtun:2005zp,Son:2007yq} conjectured a lower bound on the ratio of the shear viscosity $\eta$ and the entropy density $s$ for a general class of finite temperature field theories

\begin{equation}\label{eq:entropybound}
\frac{\eta }{ s } \geq \frac{\hbar }{ 4\pi k_{B}}.
\end{equation}

\noindent  Since all scattering channels between excitations are saturated in the quantum critical regime, the mean free time of interaction approaches the thermal equilibration time $\tau_{eq} \sim \hbar/  k_{B}T$. This gives rise to universal amplitude ratios which characterise the interacting field theory \cite{Sachdev:2011kq}. The bound \eqref{eq:entropybound} is then similar to the Mott-Ioffe-Regel limit for the minimal conductivity in disordered metals \cite{Gunnarsson:2003fk}. With the exception of particle-hole symmetric theories, however, the conductivity is determined by mechanisms which break translational symmetry. The ratio $\eta / s$ is therefore singled out as a good indicator for studying the intrinsic strength of interaction of the underlying carriers. While there are theoretical counter-examples \cite{Brigante:2008fk,Kats:2009fj} to the bound \eqref{eq:entropybound}, most phases which have been studied satisfy it \cite{Schafer:2009qq, Schafer:2007uq, Bruun:2007fk, Enss:2011gd, Muller:2009xe}.

Motivated by these developments, we consider the shear viscosity in a model of a non-Fermi liquid phase where the quasiparticle description fails. Namely, we consider the zero-gap semiconductor with a quadratic band touching and Coulomb interactions in three spatial dimensions, as first developed in \cite{Abrikosov:1971uq,Abrikosov:1971fk,  Abrikosov:1974qv}. The low energy excitations of the system are described by an interacting scale-invariant field theory, which is manifestly non-relativistic -- this fixed point has recently been dubbed the `Luttinger-Abrikosov-Beneslavskii' (LAB) phase by \cite{Moon:2013ek}. It has been suggested \cite{Moon:2013ek} that this phase could arise in strongly correlated pyrochlore iridates and might explain some of the the interesting properties of Pr$_{2}$Ir$_{2}$O$_{7}$. More generally, the model has already garnered recent attention as part of a focus on emergent quantum phases in materials with strong spin-orbit coupling and in semi-metals -- these include systems as different as graphene, the surface states of topological insulators, Weyl semimetals or nodal superconductors \cite{Vafek:2014rz,Witczak-Krempa:2014ty}.  Although there are still open question about the low energy properties of the model \cite{Herbut:2014aa}, the LAB phase might give more general insights into the nature of non-Fermi Liquid phases arising from electron-electron interactions. We briefly review the origin and properties of this model following \cite{Abrikosov:1974qv}. We shall from now on work in units where $\hbar = k_B = 1$.

\begin{figure}[tb]
   \centering
   \includegraphics[width=100pt]{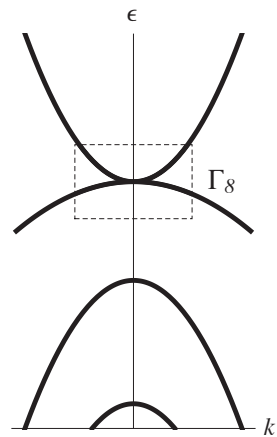} 
      \caption{Schematic of band-structure in a spin-orbit coupled s-p semiconductor with a band inversion. The quadratic band touching forms at the $\Gamma_{8}$ point; the region of the effective model giving rise to the LAB phase is shown in dashed box.}
   \label{fig:bandstruct}
\end{figure}

The bandstructure of this model is a version of the Luttinger Hamiltonian \cite{Luttinger:1956fk}, in which the four $P_{3/2}$ bands form a quadratic zero-gap semiconductor (Figure \ref{fig:bandstruct}). 
 The classic examples of this band-structure are $\alpha$-Sn and HgTe, the later has received much attention due to its importance in making topological insulators \cite{Bernevig:2006kx, Konig:2007qf}.
 The Hamiltonian for the effective model with the Fermi level at the degeneracy point and Coulomb interactions is 

\begin{equation}\label{eq:Ham}
H = \int\!\!\ud^{3} x \left[ c (\partial_{i}\psi^{\dag})A_{ij}(\partial_{j}\psi) + \frac{1}{2}  \int\!\!\ud^{3} x' \frac{\rho(\bx)\rho(\bx')}{\vert \bx-\bx' \vert}  \right]
\end{equation}

\noindent where $\psi(\bx)$ are four component electron fields, $\rho(\bx) = e \psi^{\dag}(\bx)\psi(\bx)$ is the electric charge density, and $A_{ij}$ are  matrices characterizing the bands. We take

\begin{equation}\label{eq:Amatrix}
A_{ij} = \frac{1}{2} \left\{J_{i},J_{j}\right\} - \frac{1}{3} J^{2} \delta_{ij}
\end{equation} 

\noindent where the $J_{i}$ are the usual $3/2$ angular momentum matrices. The energy spectrum of the free states has the simple form

\begin{equation}
\epsilon(\bk) =\pm ck^{2}
\end{equation}

\noindent with each band being two-fold degenerate. The $O_{h}$ symmetry in a crystal allows a more general form of the kinetic energy, which breaks both the isotropy and particle-hole symmetry of the dispersion \footnote{We note that there is no true particle-hole symmetry due to the spin structure of the system (see Supplementary Material \cite{Note3}).}, such as adding the term $\propto  (\partial_{i}\psi^{\dag})(\partial_{i}\psi)$. However, it is found that these terms are irrelevant in the renormalization-group (RG) sense \cite{Abrikosov:1974qv, Moon:2013ek}. We shall consider the low energy regime, where these terms are asymptotically small and can be neglected.

The Coulomb interaction between electrons is RG relevant in three spatial dimensions, but marginal if we consider the model generalized to four dimensions. We can therefore use an epsilon expansion ($\vare = 4-d  \ll 1$) with non-integer dimension $d$ to control the perturbative RG calculation and access the zero-temperature fixed point. Within the $\vare$-expansion, the dimensionless coupling constant $\alpha = e^2 / c \Lambda^\vare$ evolves under the renormalization group and one finds that the system flows towards a stable Wilson-Fisher fixed point \cite{Abrikosov:1974qv,Moon:2013ek}; here $\Lambda$ is the energy scale where the Hamiltonian \eqref{eq:Ham} is defined.  In particular, in the scaling limit close to the fixed point the propagator of the auxiliary Coulomb field remains unscreened and takes the form

\begin{equation}\label{eq:coulomb-crit}
V(\omega, \bq) = \frac{1}{q^{2-\eta_\varphi}} S\left(\frac{\omega}{q^z}\right).
\end{equation}

\noindent Here $z$ is the anomalous exponent describing the scaling between energy and momentum $\epsilon(\bk) = c k^{z}$ and $\eta_{\varphi}$ is the anomalous scaling dimension of the Coulomb field. The scaling function $S(x)$ is defined so that in the asymptotic limit $S \to 1$ as $x \to 0$. Using momentum shell integration we find the leading corrections $z = 2 - ({9}/{56})\vare$ and $\eta_{\varphi} = \alpha^{*}/{2\pi}$, where $\alpha^{*} = ({32 \pi}/{21})\vare$ is the fixed point coupling. We note that these values depend on the way we perform the analytic continuation of the $J=3/2$ spin structure to non-integer dimensions as discussed in detail in the Supplementary Material \footnote{See Supplementary Material for details about the renormalization-group analysis of the LAB model and technical details of the kinetic equation calculation.}. One can also develop a controlled RG calculation in $d=3$ in the limit of a large number $N$ of electronic fields \cite{Abrikosov:1974qv}, which gives a consistent picture with the fixed point found above \footnote{We will not pursue this calculation here. This restriction is due to technical feature of the kinetic equation calculation. A consistent large-$N$ theory needs to includes the full frequency dependent 1-loop renormalized interaction; see \cite{Sachdev:2011kq}.}.

We focus on studying the shear viscosity $\eta$ at finite temperatures. The shear viscosity is the transport coefficient which characterizes the relaxation of a transverse momentum gradient back to local equilibrium. Considering slow variations of the local momentum $\vect{P}(\bx)$ on a lengthscale much larger than the mean free path, the leading dissipative contribution to the  stress tensor $T_{ij}$ defines the viscosities. For a pure shear flow

\begin{equation}
\Delta T_{ij} = - \eta\left[\Pdd{x_j}{V_i}+ \Pdd{x_i}{V_j} - \frac{2}{d} \delta_{ij} (\partial_k V_k) \right],
\end{equation}

\noindent where $\vect{V} = 2c \vect{P}$. Being linear response coefficients, the viscosities can be written as Kubo formulae; for the real part of the shear viscosity \cite{[{{This form assumes the hydrodynamic limit, so the viscosity is well defined and certain $\delta(\omega)$ terms in the general Kubo formula cancel. In general, the Kubo formula for the complex viscosity also contains additional contact terms. See the in-depth discussion in }}][{}]{Bradlyn:2012uq}}

\begin{equation}\label{eq:KuboViscosity}
\eta (\omega) =  \frac{1}{{\omega} L^{d}}\cdot \Re\int_{0}^{\infty}\!\!\ud t \   {e^{i\omega^{+} t}}  \left\langle \left[ T_{xy}(t), T_{xy}(0)\right] \right\rangle
\end{equation}

\noindent where $\omega^{+} = \omega +i 0^{+}$ and $L^{d}$ the volume of the system.

To understand the transport of the LAB phase at finite temperatures, we emphasize the similarity to a system in the vicinity of a quantum critical point \cite{Sachdev:2011kq}. The long distance behaviour of the system is described by a scale invariant, interacting field theory, whose transport properties are described by universal functions. At finite frequencies there are two regimes for the transport coefficients. The collision-less limit $\omega/T \gg 1 $ is dominated by particle-hole production; the effect of interactions is small and the form of $\eta$ may be found  from the RG flow. In contrast, the limit  $\omega / T \ll 1$ is dominated by collisions between thermally excited particles and holes. In a non-interacting system, the excited particles move ballistically and give a delta-function peak; this delta function is broadened by the Coulomb interactions.   Phenomenologically, we can understand the dc transport properties from a simple mean free time argument. The temperature is the only characteristic energy scale in the problem; the characteristic length scale is  $\sim ({c/T})^{1/z}$ and the mean free path is $\ell \sim \alpha^{-2} ({c/T})^{1/z}$, which includes the dimensionless scattering rate.  A simple Drude theory would predict a dc shear viscosity $\eta \sim n \bar{v} \ell/c \sim  \alpha^{-2} (T/c)^{d/z}$. The scaling of the temperature is exactly that of the entropy density $s \sim (T/c)^{d/z}$ as expected.

We now develop a kinetic equation approach to calculate the shear viscosity to leading order in the $\vare$-expansion, which will recover the above considerations and also give a numerical estimate for the value of $\eta / s$. Formally, the kinetic equation makes use of a semi-classical expansion and assumes quasiparticles which interact locally. It was shown that in the weakly coupled finite-temperature theory of a scalar field \cite{Jeon:1995fe, Jeon:1996xr}, the diagrammatic summation in a Kubo calculation using \eqref{eq:KuboViscosity} matches the results from a high temperature effective kinetic equation. The LAB phase does not contain quasi-particles, but for small $\vare$ the quasi-particles become well defined and this approach is justified during the calculation. The long range Coulomb interaction will be screened at finite temperature on the Debye lengthscale $\ell_{D} \sim \sqrt{\alpha T/ c}$, which is much less then the mean free path, justifying the local collision term.

We write the fields  $\psi$ in terms of particle and hole eigenstates:

\begin{equation}
\psi(\bx) = \int_{\bk} \left[u_{\sigma}(\bk) c_{\sigma}(\bk) e^{i \bk\cdot \bx} + v_{\sigma}(\bk) h_{\sigma}^{\dag}(\bk) e^{-i \bk\cdot \bx} \right]
\end{equation} 

\noindent where $\sigma  =  \pm$ is the helicity of the state, $u_{\sigma}(\bk), v_{\sigma}(\bk)$ are spinor factors and $\int_{\bk} = \int (\ud^{d}k)/(2\pi)^{d}$. Since we have a four-band model, the most general semi-classical kinetic equation is for a matrix of distribution of functions including both diagonal distribution functions $ \sim \langle c^{\dag}_{\sigma} c_{\sigma}\rangle, \langle h^{\dag}_{\sigma} h_{\sigma}\rangle$ and distribution function that describe particle-hole pairs $\sim \langle c_{\sigma} h_{\sigma'}\rangle,\langle c^{\dag}_{\sigma} h^{\dag}_{\sigma'}\rangle$. Here, we neglect the latter type of distribution function, since we are  interested in the dc limit of the viscosity.  Additionally, we shall drop spin-orbit coupling in matrix elements in the equation and neglect scattering in the Mandelstam s-channel, which involves electron-hole annihilation and production. This is justified in a `leading $q$' approximation, where the dominant contribution is from particles with small momentum transfer; however, this would tend to overestimate the viscosity. In this limit, the scattering matrix elements are not affected by the spin structure of the theory. We note that at leading order in $\vare$, the details of the RG scheme do not enter the form of the Coulomb interaction and we use the lowest order expression $4\pi\alpha^* c / q^2$ as the interaction between excitations. At this order, the thermal screening of the interaction is also negligible. Unlike the bare Coulomb interaction in $d=3$, there is no low-momentum divergence of the collision integral and thermal screening is not needed as a regulator \cite{Note3}.

With the above conditions, the full quantum kinetic equation encapsulating the non-equilibrium time-evolution of the system, reduces to the usual semi-classical case

\begin{equation}\label{eq:KineticEq}
\left( \partial_{t} + {\vect{v}} \cdot \partial_{\bx} + {\vect{F}}_{ext} \cdot \partial_{\bk} \right) f^{a}(t,\bx,\bk) = - C[f]
\end{equation}

\noindent where $a$ labels both the species (electrons or holes) and helicity, $\vect{v} = 2c\bk$ is the particle velocity and $C[f]$ is the collision term describing two-particle scattering. The method of extracting the viscosity is standard -- one considers the slowly varying local momentum as an external disturbance $X_{ij} = \left[\partial_{j}{P_i}+ \partial_{i}{P_j} - ({2} \delta_{ij}/d) (\partial_k P_k) \right]/2$, assumes this weakly modifies the equilibrium distribution function $f_{0}(\bk) =1/ \left[\exp(ck^{2} / T) +1\right]$ and uses this modified distribution function to find the stress tensor $T_{ij}$ and hence $\eta$. We linearize the distribution function $f^{a}(\bk)  = f_{0}(\bk) + \delta f^{a}(\bk)$, where

\begin{equation}\label{eq:DistExpand}
\delta f^{a}(\bk) =  \beta [1-f_{0}(\bk)] f_{0}(\bk) \chi_{ij}^{a}(\bk) X_{ij}
\end{equation}

\noindent and $\chi^{a}_{ij}(\bk) = \sqrt{d/(d-1)} \left( k_{i} k_{j} - \delta_{ij}k^{2}/d\right)\chi^{a}(k)$ has the appropriate symmetry for the shear flow factored out, so that $\chi^{a}(k)$ is a function of the magnitude $k$ only.

\begin{figure}[tb]
   \centering
   \includegraphics[width=\linewidth, height=150pt]{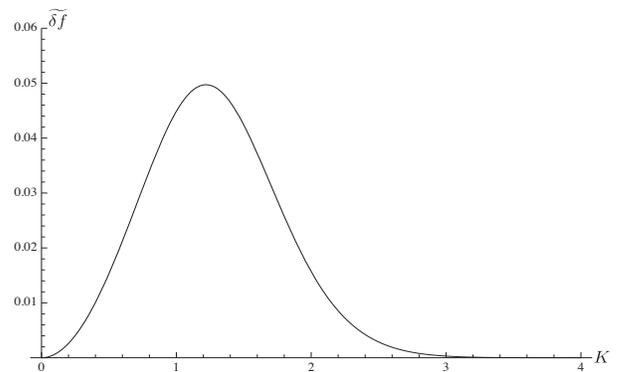} 
      \caption{Plot of optimized dimensionless function $\widetilde{\delta f}(K)= - ({T(\alpha^{*})^{2}}{c}) \cdot  \chi\left(K \right) K^{2} [1-f_{0}(K)] f_{0}(K)$ as a function of $K = k\sqrt{{T}/{c}}$.}
   \label{fig:optVisc}
\end{figure}

\noindent For the stress tensor, we consider the contributions from the diagonal fermion distribution functions

\begin{equation}
T_{ij} = \sum_{a} \int_{\bk}2c k_{i} k_{j} f^{a}_{\sigma}(\bk)
\end{equation}

\noindent In addition to neglecting the distribution function describing particle-hole creation, we neglect the contribution from the coulomb interaction between carriers to the stress tensor; we find this to be sub-leading in $\vare$ in our approach (see Supplementary Material \cite{Note3}).

We solve the kinetic equation using a variational approach  \cite{Lifshitz:1981fk, Ziman:1960rz, Arnold:2000vl}.  The kinetic equation, linearized for $\chi$, can be viewed as an operator equation on function space $\vert S_{ij} \rangle = C \vert  \chi_{ij}  \rangle$. Here $S_{ij}$ corresponds to the streaming term in \eqref{eq:KineticEq}; $C$ is an operator encoding the collision term and is Hermitian with respect to an inner product $\langle f , g \rangle =\sum_{a}\sum_{i,j} \int_{\bk}  f^{a}_{ij}(\bk) g^{a}_{ij}(\bk) $. Finding the solution $\chi(k)$ of is equivalent to maximizing the functional $Q[\chi] = \langle \chi_{ij}, S_{ij}\rangle - \langle   \chi_{ij}, C  \chi_{ij}\rangle / 2$ with respect to variations in $\chi$.  Physically, this means that the realized solution is that which maximizes entropy production subject to the constraints imposed by the external disturbance and subsequent time evolution of the system.  We project $\chi$ onto a set of 12 basis functions and  optimise the coefficients numerically to find the variational approximation (see Supplementary Material \cite{Note3});  the optimal function is shown in Figure \ref{fig:optVisc}.

Using this solution, we find the viscosity

\begin{equation}
\eta \simeq  \frac{ 3.1}{(\alpha^{*})^{2}} \left(\frac{T}{c}\right)^{2},
\end{equation}

\noindent Since we are calculating $\eta/s$ to leading order in $\vare$, we can use the entropy density for the unrenormalized band-structure in $d=4$

\begin{equation}
s = \frac{9 \zeta(3)}{16\pi^{2}}  \left(\frac{T}{c}\right)^{2},
\end{equation}

\noindent along with the fixed point value $\alpha^* = ({32 \pi}/{21})\vare$ to finally obtain

\begin{equation}\label{eq:etasfin}
\frac{\eta}{s}= \frac{0.63}{\vare^{2}}.
\end{equation}

\noindent Setting $\vare = 1$ gives $4\pi \eta/s = 8.0$; consistent with the original bound \eqref{eq:entropybound}.  A priori, there was perhaps reason to believe that this model might strongly violate the original bound, given the unusual properties of the LAB phase -- strong interactions, both particle and hole carriers, no Galilean invariance and an anomalous scaling $z$ with softer excitations than in a relativistic system. The value of $\eta/s \simeq 0.63$ is, however, similar to the values found in the unitary Fermi gas (${\eta}/{s} \leq 0.5$)  or in the quark-gluon plasma (${\eta}/{s} \leq 0.4$) \cite{Schafer:2009qq}.

The LAB phase only emerged with the chemical potential at the $\Gamma_{8}$ degeneracy point, so the system lacked a Fermi surface. Changing the chemical potential away sufficiently, we expect to recover the usual Fermi liquid behaviour. For a Fermi liquid $\eta / s \sim (\vare_{F} / T)^{3}$ is temperature dependant \cite{Abrikosov:1959uq}; in this limit the quasi-particles are well defined and weakly interacting.  By comparison, graphene with Coulomb interactions is described by a field theory with marginally irrelevant interactions, but stays in a scaling regime for a large range of energies \cite{Son:2007fv, Sheehy:2007pd}. The ratio $\eta / s \sim 1/\alpha(T)^{2}$ has a form similar to the quantum critical case \eqref{eq:etasfin}, except that the effective coupling constant $\alpha(T) \sim 1/ \log(T_{\Lambda} /T)$ ultimately renormalizes to zero as $T \to 0$ \cite{Muller:2009xe}. In both of these cases, $\eta / s$ diverges as $T \to 0$ in contrast to the universal ratio obtained for the LAB phase. 

Throughout our analysis, we have worked in the $\vare$-expansion to retain analytic control, both for the description of the RG flow and the kinetic equation calculation.  The use of the $\vare$-expansion is well established \cite{Wilson:1974aa, Zinn-Justin:2002aa}. Determining its accuracy for finite temperature transport coefficients in quantum critical systems by comparison to numerical simulations is an active research area for which we refer to the literature \footnote{See e.g. \cite{Sachdev:2011kq, Gazit:2013aa, Witczak-Krempa:2014aa} and references therein}. The extrapolation to the physical dimension is,  however, always a sensitive matter and relies on the fixed point evolving smoothly as $\vare \rightarrow 1$. This may fail if there is interference from another fixed point. While there is no other fixed point at small $\vare$, it was suggested in \cite{Herbut:2014aa} that such a fixed point might arise at strong coupling and destabilize LAB phase for the physical dimension $d=3$; this is still an open question \cite{Janssen:2015aa}.

Finally, we want to briefly comment on the experimental implications of the above discussion. In order to define the viscosity, the mean free path of effects which violate momentum conservation, such as disorder or umklapp processes, must be much larger than the mean free path of electron-electron interaction. This hydrodynamic regime is generally difficult to access in a solid state system, although, for example, recent experiments in PdCoO$_{2}$ observe very low resistivity and a temperature behaviour consistent with phonon drag \cite{Hicks:2012qf}. In this regime, we might expect the viscosity to dominate the damping of acoustic waves and also to observe viscous drag effects, which may be a route to accessing $\eta$. Alternatively, the possibility of more dramatic signatures due to electronic turbulence has been suggested in low viscosity phases \cite{Muller:2009xe}; the LAB phase would be a prime candidate for searching for such effects. Overall, we anticipate that the hydrodynamic regime will become increasingly important in the study of solid state systems.

\begin{acknowledgments}
I would like to thank S.A.~Hartnoll, D.E.~Khmelnitskii, S.A.~Parameswaran, A.C.~Potter, W.~Zwerger, and in particular A.~Vishwanath for helpful discussions and comments. Support from the National Science Foundation under grant NSF-DMR 1206728 is acknowledged.
\end{acknowledgments}

\bibliography{ZGS2-WriteUp}

\begin{thebibliography}{46}%
\makeatletter
\providecommand \@ifxundefined [1]{%
 \@ifx{#1\undefined}
}%
\providecommand \@ifnum [1]{%
 \ifnum #1\expandafter \@firstoftwo
 \else \expandafter \@secondoftwo
 \fi
}%
\providecommand \@ifx [1]{%
 \ifx #1\expandafter \@firstoftwo
 \else \expandafter \@secondoftwo
 \fi
}%
\providecommand \natexlab [1]{#1}%
\providecommand \enquote  [1]{``#1''}%
\providecommand \bibnamefont  [1]{#1}%
\providecommand \bibfnamefont [1]{#1}%
\providecommand \citenamefont [1]{#1}%
\providecommand \href@noop [0]{\@secondoftwo}%
\providecommand \href [0]{\begingroup \@sanitize@url \@href}%
\providecommand \@href[1]{\@@startlink{#1}\@@href}%
\providecommand \@@href[1]{\endgroup#1\@@endlink}%
\providecommand \@sanitize@url [0]{\catcode `\\12\catcode `\$12\catcode
  `\&12\catcode `\#12\catcode `\^12\catcode `\_12\catcode `\%12\relax}%
\providecommand \@@startlink[1]{}%
\providecommand \@@endlink[0]{}%
\providecommand \url  [0]{\begingroup\@sanitize@url \@url }%
\providecommand \@url [1]{\endgroup\@href {#1}{\urlprefix }}%
\providecommand \urlprefix  [0]{URL }%
\providecommand \Eprint [0]{\href }%
\providecommand \doibase [0]{http://dx.doi.org/}%
\providecommand \selectlanguage [0]{\@gobble}%
\providecommand \bibinfo  [0]{\@secondoftwo}%
\providecommand \bibfield  [0]{\@secondoftwo}%
\providecommand \translation [1]{[#1]}%
\providecommand \BibitemOpen [0]{}%
\providecommand \bibitemStop [0]{}%
\providecommand \bibitemNoStop [0]{.\EOS\space}%
\providecommand \EOS [0]{\spacefactor3000\relax}%
\providecommand \BibitemShut  [1]{\csname bibitem#1\endcsname}%
\let\auto@bib@innerbib\@empty
\bibitem [{\citenamefont {Sachdev}(2011)}]{Sachdev:2011kq}%
  \BibitemOpen
  \bibfield  {author} {\bibinfo {author} {\bibfnamefont {S.}~\bibnamefont
  {Sachdev}},\ }\href@noop {} {\emph {\bibinfo {title} {Quantum Phase
  Transitions}}},\ \bibinfo {edition} {2nd}\ ed.\ (\bibinfo  {publisher}
  {Cambridge University Press},\ \bibinfo {address} {Cambridge},\ \bibinfo
  {year} {2011})\BibitemShut {NoStop}%
\bibitem [{\citenamefont {Sachdev}(2012)}]{Sachdev:2012sf}%
  \BibitemOpen
  \bibfield  {author} {\bibinfo {author} {\bibfnamefont {S.}~\bibnamefont
  {Sachdev}},\ }\href {\doibase 10.1146/annurev-conmatphys-020911-125141}
  {\bibfield  {journal} {\bibinfo  {journal} {Annu. Rev. Con. Mat. Phys.}\
  }\textbf {\bibinfo {volume} {3}},\ \bibinfo {pages} {9} (\bibinfo {year}
  {2012})}\BibitemShut {NoStop}%
\bibitem [{\citenamefont {Hartnoll}(2009)}]{Hartnoll:2009xy}%
  \BibitemOpen
  \bibfield  {author} {\bibinfo {author} {\bibfnamefont {S.~A.}\ \bibnamefont
  {Hartnoll}},\ }\href {http://stacks.iop.org/0264-9381/26/i=22/a=224002}
  {\bibfield  {journal} {\bibinfo  {journal} {Classical and Quantum Gravity}\
  }\textbf {\bibinfo {volume} {26}},\ \bibinfo {pages} {224002} (\bibinfo
  {year} {2009})}\BibitemShut {NoStop}%
\bibitem [{\citenamefont {Kovtun}\ \emph {et~al.}(2005)\citenamefont {Kovtun},
  \citenamefont {Son},\ and\ \citenamefont {Starinets}}]{Kovtun:2005zp}%
  \BibitemOpen
  \bibfield  {author} {\bibinfo {author} {\bibfnamefont {P.~K.}\ \bibnamefont
  {Kovtun}}, \bibinfo {author} {\bibfnamefont {D.~T.}\ \bibnamefont {Son}}, \
  and\ \bibinfo {author} {\bibfnamefont {A.~O.}\ \bibnamefont {Starinets}},\
  }\href@noop {} {\bibfield  {journal} {\bibinfo  {journal} {Phys. Rev. Lett.}\
  }\textbf {\bibinfo {volume} {94}},\ \bibinfo {pages} {111601} (\bibinfo
  {year} {2005})}\BibitemShut {NoStop}%
\bibitem [{\citenamefont {Son}\ and\ \citenamefont
  {Starinets}(2007)}]{Son:2007yq}%
  \BibitemOpen
  \bibfield  {author} {\bibinfo {author} {\bibfnamefont {D.~T.}\ \bibnamefont
  {Son}}\ and\ \bibinfo {author} {\bibfnamefont {A.~O.}\ \bibnamefont
  {Starinets}},\ }\href {\doibase 10.1146/annurev.nucl.57.090506.123120}
  {\bibfield  {journal} {\bibinfo  {journal} {Ann. Rev. Nucl. Part. Sci.}\
  }\textbf {\bibinfo {volume} {57}},\ \bibinfo {pages} {95} (\bibinfo {year}
  {2007})}\BibitemShut {NoStop}%
\bibitem [{\citenamefont {Gunnarsson}\ \emph {et~al.}(2003)\citenamefont
  {Gunnarsson}, \citenamefont {Calandra},\ and\ \citenamefont
  {Han}}]{Gunnarsson:2003fk}%
  \BibitemOpen
  \bibfield  {author} {\bibinfo {author} {\bibfnamefont {O.}~\bibnamefont
  {Gunnarsson}}, \bibinfo {author} {\bibfnamefont {M.}~\bibnamefont
  {Calandra}}, \ and\ \bibinfo {author} {\bibfnamefont {J.~E.}\ \bibnamefont
  {Han}},\ }\href {\doibase 10.1103/RevModPhys.75.1085} {\bibfield  {journal}
  {\bibinfo  {journal} {Rev. Mod. Phys.}\ }\textbf {\bibinfo {volume} {75}},\
  \bibinfo {pages} {1085} (\bibinfo {year} {2003})}\BibitemShut {NoStop}%
\bibitem [{\citenamefont {Brigante}\ \emph {et~al.}(2008)\citenamefont
  {Brigante}, \citenamefont {Liu}, \citenamefont {Myers}, \citenamefont
  {Shenker},\ and\ \citenamefont {Yaida}}]{Brigante:2008fk}%
  \BibitemOpen
  \bibfield  {author} {\bibinfo {author} {\bibfnamefont {M.}~\bibnamefont
  {Brigante}}, \bibinfo {author} {\bibfnamefont {H.}~\bibnamefont {Liu}},
  \bibinfo {author} {\bibfnamefont {R.~C.}\ \bibnamefont {Myers}}, \bibinfo
  {author} {\bibfnamefont {S.}~\bibnamefont {Shenker}}, \ and\ \bibinfo
  {author} {\bibfnamefont {S.}~\bibnamefont {Yaida}},\ }\href {\doibase
  10.1103/PhysRevLett.100.191601} {\bibfield  {journal} {\bibinfo  {journal}
  {Phys. Rev. Lett.}\ }\textbf {\bibinfo {volume} {100}},\ \bibinfo {pages}
  {191601} (\bibinfo {year} {2008})}\BibitemShut {NoStop}%
\bibitem [{\citenamefont {Kats}\ and\ \citenamefont
  {Petrov}(2009)}]{Kats:2009fj}%
  \BibitemOpen
  \bibfield  {author} {\bibinfo {author} {\bibfnamefont {Y.}~\bibnamefont
  {Kats}}\ and\ \bibinfo {author} {\bibfnamefont {P.}~\bibnamefont {Petrov}},\
  }\href {http://stacks.iop.org/1126-6708/2009/i=01/a=044} {\bibfield
  {journal} {\bibinfo  {journal} {J. High Energy Phys.}\ }\textbf {\bibinfo
  {volume} {2009}},\ \bibinfo {pages} {044} (\bibinfo {year}
  {2009})}\BibitemShut {NoStop}%
\bibitem [{\citenamefont {Sch{\"a}fer}\ and\ \citenamefont
  {Teaney}(2009)}]{Schafer:2009qq}%
  \BibitemOpen
  \bibfield  {author} {\bibinfo {author} {\bibfnamefont {T.}~\bibnamefont
  {Sch{\"a}fer}}\ and\ \bibinfo {author} {\bibfnamefont {D.}~\bibnamefont
  {Teaney}},\ }\href@noop {} {\bibfield  {journal} {\bibinfo  {journal} {Rep.
  Prog. Phys.}\ }\textbf {\bibinfo {volume} {72}},\ \bibinfo {pages} {126001}
  (\bibinfo {year} {2009})}\BibitemShut {NoStop}%
\bibitem [{\citenamefont {Sch\"afer}(2007)}]{Schafer:2007uq}%
  \BibitemOpen
  \bibfield  {author} {\bibinfo {author} {\bibfnamefont {T.}~\bibnamefont
  {Sch\"afer}},\ }\href {\doibase 10.1103/PhysRevA.76.063618} {\bibfield
  {journal} {\bibinfo  {journal} {Phys. Rev. A}\ }\textbf {\bibinfo {volume}
  {76}},\ \bibinfo {pages} {063618} (\bibinfo {year} {2007})}\BibitemShut
  {NoStop}%
\bibitem [{\citenamefont {Bruun}\ and\ \citenamefont
  {Smith}(2007)}]{Bruun:2007fk}%
  \BibitemOpen
  \bibfield  {author} {\bibinfo {author} {\bibfnamefont {G.~M.}\ \bibnamefont
  {Bruun}}\ and\ \bibinfo {author} {\bibfnamefont {H.}~\bibnamefont {Smith}},\
  }\href {\doibase 10.1103/PhysRevA.75.043612} {\bibfield  {journal} {\bibinfo
  {journal} {Phys. Rev. A}\ }\textbf {\bibinfo {volume} {75}},\ \bibinfo
  {pages} {043612} (\bibinfo {year} {2007})}\BibitemShut {NoStop}%
\bibitem [{\citenamefont {Enss}\ \emph {et~al.}(2011)\citenamefont {Enss},
  \citenamefont {Haussmann},\ and\ \citenamefont {Zwerger}}]{Enss:2011gd}%
  \BibitemOpen
  \bibfield  {author} {\bibinfo {author} {\bibfnamefont {T.}~\bibnamefont
  {Enss}}, \bibinfo {author} {\bibfnamefont {R.}~\bibnamefont {Haussmann}}, \
  and\ \bibinfo {author} {\bibfnamefont {W.}~\bibnamefont {Zwerger}},\
  }\href@noop {} {\bibfield  {journal} {\bibinfo  {journal} {Ann. Phys. (NY)}\
  }\textbf {\bibinfo {volume} {326}},\ \bibinfo {pages} {770 } (\bibinfo {year}
  {2011})}\BibitemShut {NoStop}%
\bibitem [{\citenamefont {M{\"u}ller}\ \emph {et~al.}(2009)\citenamefont
  {M{\"u}ller}, \citenamefont {Schmalian},\ and\ \citenamefont
  {Fritz}}]{Muller:2009xe}%
  \BibitemOpen
  \bibfield  {author} {\bibinfo {author} {\bibfnamefont {M.}~\bibnamefont
  {M{\"u}ller}}, \bibinfo {author} {\bibfnamefont {J.}~\bibnamefont
  {Schmalian}}, \ and\ \bibinfo {author} {\bibfnamefont {L.}~\bibnamefont
  {Fritz}},\ }\href@noop {} {\bibfield  {journal} {\bibinfo  {journal} {Phys.
  Rev. Lett.}\ }\textbf {\bibinfo {volume} {103}},\ \bibinfo {pages} {025301}
  (\bibinfo {year} {2009})}\BibitemShut {NoStop}%
\bibitem [{\citenamefont {{Abrikosov}}\ and\ \citenamefont
  {{Beneslavskii}}(1971{\natexlab{a}})}]{Abrikosov:1971uq}%
  \BibitemOpen
  \bibfield  {author} {\bibinfo {author} {\bibfnamefont {A.~A.}\ \bibnamefont
  {{Abrikosov}}}\ and\ \bibinfo {author} {\bibfnamefont {S.~D.}\ \bibnamefont
  {{Beneslavskii}}},\ }\href@noop {} {\bibfield  {journal} {\bibinfo  {journal}
  {Sov.~Phys.~JETP}\ }\textbf {\bibinfo {volume} {32}},\ \bibinfo {pages} {699}
  (\bibinfo {year} {1971}{\natexlab{a}})}\BibitemShut {NoStop}%
\bibitem [{\citenamefont {{Abrikosov}}\ and\ \citenamefont
  {{Beneslavskii}}(1971{\natexlab{b}})}]{Abrikosov:1971fk}%
  \BibitemOpen
  \bibfield  {author} {\bibinfo {author} {\bibfnamefont {A.~A.}\ \bibnamefont
  {{Abrikosov}}}\ and\ \bibinfo {author} {\bibfnamefont {S.~D.}\ \bibnamefont
  {{Beneslavskii}}},\ }\href@noop {} {\bibfield  {journal} {\bibinfo  {journal}
  {J.~Low. Temp.~Phys.}\ }\textbf {\bibinfo {volume} {5}},\ \bibinfo {pages}
  {141} (\bibinfo {year} {1971}{\natexlab{b}})}\BibitemShut {NoStop}%
\bibitem [{\citenamefont {{Abrikosov}}(1974)}]{Abrikosov:1974qv}%
  \BibitemOpen
  \bibfield  {author} {\bibinfo {author} {\bibfnamefont {A.~A.}\ \bibnamefont
  {{Abrikosov}}},\ }\href@noop {} {\bibfield  {journal} {\bibinfo  {journal}
  {Sov.~Phys.~JETP}\ }\textbf {\bibinfo {volume} {39}},\ \bibinfo {pages} {709}
  (\bibinfo {year} {1974})}\BibitemShut {NoStop}%
\bibitem [{\citenamefont {Moon}\ \emph {et~al.}(2013)\citenamefont {Moon},
  \citenamefont {Xu}, \citenamefont {Kim},\ and\ \citenamefont
  {Balents}}]{Moon:2013ek}%
  \BibitemOpen
  \bibfield  {author} {\bibinfo {author} {\bibfnamefont {E.-G.}\ \bibnamefont
  {Moon}}, \bibinfo {author} {\bibfnamefont {C.}~\bibnamefont {Xu}}, \bibinfo
  {author} {\bibfnamefont {Y.~B.}\ \bibnamefont {Kim}}, \ and\ \bibinfo
  {author} {\bibfnamefont {L.}~\bibnamefont {Balents}},\ }\href@noop {}
  {\bibfield  {journal} {\bibinfo  {journal} {Phys. Rev. Lett.}\ }\textbf
  {\bibinfo {volume} {111}},\ \bibinfo {pages} {206401} (\bibinfo {year}
  {2013})}\BibitemShut {NoStop}%
\bibitem [{\citenamefont {Vafek}\ and\ \citenamefont
  {Vishwanath}(2014)}]{Vafek:2014rz}%
  \BibitemOpen
  \bibfield  {author} {\bibinfo {author} {\bibfnamefont {O.}~\bibnamefont
  {Vafek}}\ and\ \bibinfo {author} {\bibfnamefont {A.}~\bibnamefont
  {Vishwanath}},\ }\href {\doibase 10.1146/annurev-conmatphys-031113-133841}
  {\bibfield  {journal} {\bibinfo  {journal} {Annu. Rev. Con. Mat. Phys.}\
  }\textbf {\bibinfo {volume} {5}},\ \bibinfo {pages} {83} (\bibinfo {year}
  {2014})}\BibitemShut {NoStop}%
\bibitem [{\citenamefont {Witczak-Krempa}\ \emph
  {et~al.}(2014{\natexlab{a}})\citenamefont {Witczak-Krempa}, \citenamefont
  {Chen}, \citenamefont {Kim},\ and\ \citenamefont
  {Balents}}]{Witczak-Krempa:2014ty}%
  \BibitemOpen
  \bibfield  {author} {\bibinfo {author} {\bibfnamefont {W.}~\bibnamefont
  {Witczak-Krempa}}, \bibinfo {author} {\bibfnamefont {G.}~\bibnamefont
  {Chen}}, \bibinfo {author} {\bibfnamefont {Y.~B.}\ \bibnamefont {Kim}}, \
  and\ \bibinfo {author} {\bibfnamefont {L.}~\bibnamefont {Balents}},\
  }\href@noop {} {\bibfield  {journal} {\bibinfo  {journal} {Annu. Rev. Con.
  Mat. Phys.}\ }\textbf {\bibinfo {volume} {5}},\ \bibinfo {pages} {57}
  (\bibinfo {year} {2014}{\natexlab{a}})}\BibitemShut {NoStop}%
\bibitem [{\citenamefont {Herbut}\ and\ \citenamefont
  {Janssen}(2014)}]{Herbut:2014aa}%
  \BibitemOpen
  \bibfield  {author} {\bibinfo {author} {\bibfnamefont {I.~F.}\ \bibnamefont
  {Herbut}}\ and\ \bibinfo {author} {\bibfnamefont {L.}~\bibnamefont
  {Janssen}},\ }\href {\doibase 10.1103/PhysRevLett.113.106401} {\bibfield
  {journal} {\bibinfo  {journal} {Phys. Rev. Lett.}\ }\textbf {\bibinfo
  {volume} {113}},\ \bibinfo {pages} {106401} (\bibinfo {year}
  {2014})}\BibitemShut {NoStop}%
\bibitem [{\citenamefont {Luttinger}(1956)}]{Luttinger:1956fk}%
  \BibitemOpen
  \bibfield  {author} {\bibinfo {author} {\bibfnamefont {J.~M.}\ \bibnamefont
  {Luttinger}},\ }\href@noop {} {\bibfield  {journal} {\bibinfo  {journal}
  {Phys. Rev.}\ }\textbf {\bibinfo {volume} {102}},\ \bibinfo {pages} {1030}
  (\bibinfo {year} {1956})}\BibitemShut {NoStop}%
\bibitem [{\citenamefont {Bernevig}\ \emph {et~al.}(2006)\citenamefont
  {Bernevig}, \citenamefont {Hughes},\ and\ \citenamefont
  {Zhang}}]{Bernevig:2006kx}%
  \BibitemOpen
  \bibfield  {author} {\bibinfo {author} {\bibfnamefont {B.~A.}\ \bibnamefont
  {Bernevig}}, \bibinfo {author} {\bibfnamefont {T.~L.}\ \bibnamefont
  {Hughes}}, \ and\ \bibinfo {author} {\bibfnamefont {S.-C.}\ \bibnamefont
  {Zhang}},\ }\href@noop {} {\bibfield  {journal} {\bibinfo  {journal}
  {Science}\ }\textbf {\bibinfo {volume} {314}},\ \bibinfo {pages} {1757}
  (\bibinfo {year} {2006})}\BibitemShut {NoStop}%
\bibitem [{\citenamefont {K{\"o}nig}\ \emph {et~al.}(2007)\citenamefont
  {K{\"o}nig}, \citenamefont {Wiedmann}, \citenamefont {Br{\"u}ne},
  \citenamefont {Roth}, \citenamefont {Buhmann}, \citenamefont {Molenkamp},
  \citenamefont {Qi},\ and\ \citenamefont {Zhang}}]{Konig:2007qf}%
  \BibitemOpen
  \bibfield  {author} {\bibinfo {author} {\bibfnamefont {M.}~\bibnamefont
  {K{\"o}nig}}, \bibinfo {author} {\bibfnamefont {S.}~\bibnamefont {Wiedmann}},
  \bibinfo {author} {\bibfnamefont {C.}~\bibnamefont {Br{\"u}ne}}, \bibinfo
  {author} {\bibfnamefont {A.}~\bibnamefont {Roth}}, \bibinfo {author}
  {\bibfnamefont {H.}~\bibnamefont {Buhmann}}, \bibinfo {author} {\bibfnamefont
  {L.~W.}\ \bibnamefont {Molenkamp}}, \bibinfo {author} {\bibfnamefont {X.-L.}\
  \bibnamefont {Qi}}, \ and\ \bibinfo {author} {\bibfnamefont {S.-C.}\
  \bibnamefont {Zhang}},\ }\href@noop {} {\bibfield  {journal} {\bibinfo
  {journal} {Science}\ }\textbf {\bibinfo {volume} {318}},\ \bibinfo {pages}
  {766} (\bibinfo {year} {2007})}\BibitemShut {NoStop}%
\bibitem [{Note1()}]{Note1}%
  \BibitemOpen
  \bibinfo {note} {We note that there is no true particle-hole symmetry due to
  the spin structure of the system (see Supplementary Material \cite
  {Note3}).}\BibitemShut {Stop}%
\bibitem [{Note2()}]{Note2}%
  \BibitemOpen
  \bibinfo {note} {See Supplementary Material for details about the
  renormalization-group analysis of the LAB model and technical details of the
  kinetic equation calculation.}\BibitemShut {Stop}%
\bibitem [{Note3()}]{Note3}%
  \BibitemOpen
  \bibinfo {note} {We will not pursue this calculation here. This restriction
  is due to technical feature of the kinetic equation calculation. A consistent
  large-$N$ theory needs to includes the full frequency dependent 1-loop
  renormalized interaction; see \cite {Sachdev:2011kq}.}\BibitemShut {Stop}%
\bibitem [{\citenamefont {Bradlyn}\ \emph {et~al.}(2012)\citenamefont
  {Bradlyn}, \citenamefont {Goldstein},\ and\ \citenamefont
  {Read}}]{Bradlyn:2012uq}%
  \BibitemOpen
  \bibfield  {author} {\bibinfo {author} {\bibfnamefont {B.}~\bibnamefont
  {Bradlyn}}, \bibinfo {author} {\bibfnamefont {M.}~\bibnamefont {Goldstein}},
  \ and\ \bibinfo {author} {\bibfnamefont {N.}~\bibnamefont {Read}},\
  }\href@noop {} {\bibfield  {journal} {\bibinfo  {journal} {Phys. Rev. B}\
  }\textbf {\bibinfo {volume} {86}},\ \bibinfo {pages} {245309} (\bibinfo
  {year} {2012})}\BibitemShut {NoStop}%
\bibitem [{\citenamefont {Jeon}(1995)}]{Jeon:1995fe}%
  \BibitemOpen
  \bibfield  {author} {\bibinfo {author} {\bibfnamefont {S.}~\bibnamefont
  {Jeon}},\ }\href@noop {} {\bibfield  {journal} {\bibinfo  {journal} {Phys.
  Rev. D}\ }\textbf {\bibinfo {volume} {52}},\ \bibinfo {pages} {3591}
  (\bibinfo {year} {1995})}\BibitemShut {NoStop}%
\bibitem [{\citenamefont {Jeon}\ and\ \citenamefont
  {Yaffe}(1996)}]{Jeon:1996xr}%
  \BibitemOpen
  \bibfield  {author} {\bibinfo {author} {\bibfnamefont {S.}~\bibnamefont
  {Jeon}}\ and\ \bibinfo {author} {\bibfnamefont {L.~G.}\ \bibnamefont
  {Yaffe}},\ }\href@noop {} {\bibfield  {journal} {\bibinfo  {journal} {Phys.
  Rev. D}\ }\textbf {\bibinfo {volume} {53}},\ \bibinfo {pages} {5799}
  (\bibinfo {year} {1996})}\BibitemShut {NoStop}%
\bibitem [{\citenamefont {{Lifshitz}}\ and\ \citenamefont
  {{Pitaevskii}}(1981)}]{Lifshitz:1981fk}%
  \BibitemOpen
  \bibfield  {author} {\bibinfo {author} {\bibfnamefont {E.~M.}\ \bibnamefont
  {{Lifshitz}}}\ and\ \bibinfo {author} {\bibfnamefont {L.~P.}\ \bibnamefont
  {{Pitaevskii}}},\ }\href@noop {} {\emph {\bibinfo {title} {Physical
  Kinetics}}},\ \bibinfo {series} {Course of Theoretical Physics},
  Vol.~\bibinfo {volume} {10}\ (\bibinfo  {publisher} {Pergamon Press},\
  \bibinfo {address} {Oxford},\ \bibinfo {year} {1981})\BibitemShut {NoStop}%
\bibitem [{\citenamefont {Ziman}(1960)}]{Ziman:1960rz}%
  \BibitemOpen
  \bibfield  {author} {\bibinfo {author} {\bibfnamefont {J.~M.}\ \bibnamefont
  {Ziman}},\ }\href@noop {} {\emph {\bibinfo {title} {Electrons and Phonons}}}\
  (\bibinfo  {publisher} {Oxford University Press},\ \bibinfo {address}
  {Oxford},\ \bibinfo {year} {1960})\BibitemShut {NoStop}%
\bibitem [{\citenamefont {Arnold}\ \emph {et~al.}(2000)\citenamefont {Arnold},
  \citenamefont {Moore},\ and\ \citenamefont {Yaffe}}]{Arnold:2000vl}%
  \BibitemOpen
  \bibfield  {author} {\bibinfo {author} {\bibfnamefont {P.}~\bibnamefont
  {Arnold}}, \bibinfo {author} {\bibfnamefont {G.~D.}\ \bibnamefont {Moore}}, \
  and\ \bibinfo {author} {\bibfnamefont {L.~G.}\ \bibnamefont {Yaffe}},\
  }\href@noop {} {\bibfield  {journal} {\bibinfo  {journal} {J. High Energy
  Phys.}\ }\textbf {\bibinfo {volume} {2000}},\ \bibinfo {pages} {001}
  (\bibinfo {year} {2000})}\BibitemShut {NoStop}%
\bibitem [{\citenamefont {Abrikosov}\ and\ \citenamefont
  {Khalatnikov}(1959)}]{Abrikosov:1959uq}%
  \BibitemOpen
  \bibfield  {author} {\bibinfo {author} {\bibfnamefont {A.~A.}\ \bibnamefont
  {Abrikosov}}\ and\ \bibinfo {author} {\bibfnamefont {I.~M.}\ \bibnamefont
  {Khalatnikov}},\ }\href@noop {} {\bibfield  {journal} {\bibinfo  {journal}
  {Reports on Progress in Physics}\ }\textbf {\bibinfo {volume} {22}},\
  \bibinfo {pages} {329} (\bibinfo {year} {1959})}\BibitemShut {NoStop}%
\bibitem [{\citenamefont {Son}(2007)}]{Son:2007fv}%
  \BibitemOpen
  \bibfield  {author} {\bibinfo {author} {\bibfnamefont {D.~T.}\ \bibnamefont
  {Son}},\ }\href {\doibase 10.1103/PhysRevB.75.235423} {\bibfield  {journal}
  {\bibinfo  {journal} {Phys. Rev. B}\ }\textbf {\bibinfo {volume} {75}},\
  \bibinfo {pages} {235423} (\bibinfo {year} {2007})}\BibitemShut {NoStop}%
\bibitem [{\citenamefont {Sheehy}\ and\ \citenamefont
  {Schmalian}(2007)}]{Sheehy:2007pd}%
  \BibitemOpen
  \bibfield  {author} {\bibinfo {author} {\bibfnamefont {D.~E.}\ \bibnamefont
  {Sheehy}}\ and\ \bibinfo {author} {\bibfnamefont {J.}~\bibnamefont
  {Schmalian}},\ }\href {\doibase 10.1103/PhysRevLett.99.226803} {\bibfield
  {journal} {\bibinfo  {journal} {Phys. Rev. Lett.}\ }\textbf {\bibinfo
  {volume} {99}},\ \bibinfo {pages} {226803} (\bibinfo {year}
  {2007})}\BibitemShut {NoStop}%
\bibitem [{\citenamefont {Wilson}\ and\ \citenamefont
  {Kogut}(1974)}]{Wilson:1974aa}%
  \BibitemOpen
  \bibfield  {author} {\bibinfo {author} {\bibfnamefont {K.~G.}\ \bibnamefont
  {Wilson}}\ and\ \bibinfo {author} {\bibfnamefont {J.}~\bibnamefont {Kogut}},\
  }\href {\doibase http://dx.doi.org/10.1016/0370-1573(74)90023-4} {\bibfield
  {journal} {\bibinfo  {journal} {Physics Reports}\ }\textbf {\bibinfo {volume}
  {12}},\ \bibinfo {pages} {75 } (\bibinfo {year} {1974})}\BibitemShut
  {NoStop}%
\bibitem [{\citenamefont {Zinn-Justin}(2002)}]{Zinn-Justin:2002aa}%
  \BibitemOpen
  \bibfield  {author} {\bibinfo {author} {\bibfnamefont {J.}~\bibnamefont
  {Zinn-Justin}},\ }\href
  {http://www.loc.gov/catdir/enhancements/fy0613/2002034631-d.html} {\emph
  {\bibinfo {title} {Quantum field theory and critical phenomena}}},\ \bibinfo
  {edition} {4th}\ ed.,\ \bibinfo {series} {International series of monographs
  on physics}, Vol.\ \bibinfo {volume} {113}\ (\bibinfo  {publisher} {Clarendon
  Press},\ \bibinfo {address} {Oxford},\ \bibinfo {year} {2002})\BibitemShut
  {NoStop}%
\bibitem [{Note4()}]{Note4}%
  \BibitemOpen
  \bibinfo {note} {See e.g. \cite {Sachdev:2011kq, Gazit:2013aa,
  Witczak-Krempa:2014aa} and references therein}\BibitemShut {NoStop}%
\bibitem [{\citenamefont {{Janssen}}\ and\ \citenamefont
  {{Herbut}}(2015)}]{Janssen:2015aa}%
  \BibitemOpen
  \bibfield  {author} {\bibinfo {author} {\bibfnamefont {L.}~\bibnamefont
  {{Janssen}}}\ and\ \bibinfo {author} {\bibfnamefont {I.~F.}\ \bibnamefont
  {{Herbut}}},\ }\href@noop {} {\bibfield  {journal} {\bibinfo  {journal}
  {ArXiv e-prints}\ } (\bibinfo {year} {2015})},\ \Eprint
  {http://arxiv.org/abs/1503.04242} {arXiv:1503.04242 [cond-mat.str-el]}
  \BibitemShut {NoStop}%
\bibitem [{\citenamefont {Hicks}\ \emph {et~al.}(2012)\citenamefont {Hicks},
  \citenamefont {Gibbs}, \citenamefont {Mackenzie}, \citenamefont {Takatsu},
  \citenamefont {Maeno},\ and\ \citenamefont {Yelland}}]{Hicks:2012qf}%
  \BibitemOpen
  \bibfield  {author} {\bibinfo {author} {\bibfnamefont {C.~W.}\ \bibnamefont
  {Hicks}}, \bibinfo {author} {\bibfnamefont {A.~S.}\ \bibnamefont {Gibbs}},
  \bibinfo {author} {\bibfnamefont {A.~P.}\ \bibnamefont {Mackenzie}}, \bibinfo
  {author} {\bibfnamefont {H.}~\bibnamefont {Takatsu}}, \bibinfo {author}
  {\bibfnamefont {Y.}~\bibnamefont {Maeno}}, \ and\ \bibinfo {author}
  {\bibfnamefont {E.~A.}\ \bibnamefont {Yelland}},\ }\href {\doibase
  10.1103/PhysRevLett.109.116401} {\bibfield  {journal} {\bibinfo  {journal}
  {Phys. Rev. Lett.}\ }\textbf {\bibinfo {volume} {109}},\ \bibinfo {pages}
  {116401} (\bibinfo {year} {2012})}\BibitemShut {NoStop}%
\bibitem [{\citenamefont {Gazit}\ \emph {et~al.}(2013)\citenamefont {Gazit},
  \citenamefont {Podolsky}, \citenamefont {Auerbach},\ and\ \citenamefont
  {Arovas}}]{Gazit:2013aa}%
  \BibitemOpen
  \bibfield  {author} {\bibinfo {author} {\bibfnamefont {S.}~\bibnamefont
  {Gazit}}, \bibinfo {author} {\bibfnamefont {D.}~\bibnamefont {Podolsky}},
  \bibinfo {author} {\bibfnamefont {A.}~\bibnamefont {Auerbach}}, \ and\
  \bibinfo {author} {\bibfnamefont {D.~P.}\ \bibnamefont {Arovas}},\ }\href
  {\doibase 10.1103/PhysRevB.88.235108} {\bibfield  {journal} {\bibinfo
  {journal} {Phys. Rev. B}\ }\textbf {\bibinfo {volume} {88}},\ \bibinfo
  {pages} {235108} (\bibinfo {year} {2013})}\BibitemShut {NoStop}%
\bibitem [{\citenamefont {Witczak-Krempa}\ \emph
  {et~al.}(2014{\natexlab{b}})\citenamefont {Witczak-Krempa}, \citenamefont
  {Sorensen},\ and\ \citenamefont {Sachdev}}]{Witczak-Krempa:2014aa}%
  \BibitemOpen
  \bibfield  {author} {\bibinfo {author} {\bibfnamefont {W.}~\bibnamefont
  {Witczak-Krempa}}, \bibinfo {author} {\bibfnamefont {E.~S.}\ \bibnamefont
  {Sorensen}}, \ and\ \bibinfo {author} {\bibfnamefont {S.}~\bibnamefont
  {Sachdev}},\ }\href {http://dx.doi.org/10.1038/nphys2913} {\bibfield
  {journal} {\bibinfo  {journal} {Nat Phys}\ }\textbf {\bibinfo {volume}
  {10}},\ \bibinfo {pages} {361} (\bibinfo {year}
  {2014}{\natexlab{b}})}\BibitemShut {NoStop}%
\bibitem [{\citenamefont {Fu}(2011)}]{Fu:2011fk}%
  \BibitemOpen
  \bibfield  {author} {\bibinfo {author} {\bibfnamefont {L.}~\bibnamefont
  {Fu}},\ }\href {\doibase 10.1103/PhysRevLett.106.106802} {\bibfield
  {journal} {\bibinfo  {journal} {Phys. Rev. Lett.}\ }\textbf {\bibinfo
  {volume} {106}},\ \bibinfo {pages} {106802} (\bibinfo {year}
  {2011})}\BibitemShut {NoStop}%
\bibitem [{\citenamefont {Murakami}\ \emph {et~al.}(2004)\citenamefont
  {Murakami}, \citenamefont {Nagaosa},\ and\ \citenamefont
  {Zhang}}]{Murakami:2004qf}%
  \BibitemOpen
  \bibfield  {author} {\bibinfo {author} {\bibfnamefont {S.}~\bibnamefont
  {Murakami}}, \bibinfo {author} {\bibfnamefont {N.}~\bibnamefont {Nagaosa}}, \
  and\ \bibinfo {author} {\bibfnamefont {S.-C.}\ \bibnamefont {Zhang}},\ }\href
  {\doibase 10.1103/PhysRevB.69.235206} {\bibfield  {journal} {\bibinfo
  {journal} {Phys. Rev. B}\ }\textbf {\bibinfo {volume} {69}},\ \bibinfo
  {pages} {235206} (\bibinfo {year} {2004})}\BibitemShut {NoStop}%
\bibitem [{\citenamefont {Hahn}(2005)}]{Hahn:2005fk}%
  \BibitemOpen
  \bibfield  {author} {\bibinfo {author} {\bibfnamefont {T.}~\bibnamefont
  {Hahn}},\ }\href@noop {} {\bibfield  {journal} {\bibinfo  {journal} {Computer
  Physics Communications}\ }\textbf {\bibinfo {volume} {168}},\ \bibinfo
  {pages} {78 } (\bibinfo {year} {2005})}\BibitemShut {NoStop}%
\bibitem [{\citenamefont {Altland}\ and\ \citenamefont
  {Simons}(2010)}]{Altland:2010uq}%
  \BibitemOpen
  \bibfield  {author} {\bibinfo {author} {\bibfnamefont {A.}~\bibnamefont
  {Altland}}\ and\ \bibinfo {author} {\bibfnamefont {B.~D.}\ \bibnamefont
  {Simons}},\ }\href@noop {} {\emph {\bibinfo {title} {Condensed Matter Field
  Theory}}},\ \bibinfo {edition} {2nd}\ ed.\ (\bibinfo  {publisher} {Cambridge
  University Press},\ \bibinfo {address} {Cambridge},\ \bibinfo {year}
  {2010})\BibitemShut {NoStop}%
\end{thebibliography}%

\onecolumngrid
\newpage


\appendix
\section*{Supplementary material}

\subsection{Model and Renormalization Group Analysis}

\subsubsection{Spin-Orbit Coupled Bandstructure}

The group theoretic structure of the Luttinger Hamiltonian is well established \cite{Luttinger:1956fk}. The four states at the degneracy point belong to the $J={3}/{2}$ double valued representation of the octahedral group $O_{h}$ ($\Gamma_{8}$).  For the model which is isotropic in all directions we took

\begin{equation}\label{eq:AmatrixApp}
A_{ij} = \frac{1}{2}\{J_{i},J_{j}\}-\frac{1}{3}J^{2}\delta_{ij} 
\end{equation}

\noindent in the physical dimension $d=3$. There is an ambiguity in how to extend this structure to $d=4-\vare$ needed for the RG calculation. The choice will affect both the fixed point value of the interaction found from the RG analysis as well as the scattering elements entering the kinetic equation. The simplest choice is to formally use the $d=3$ structure.  One can also use a spin structure in $d=4$ -- the starting point of the $\vare$-expansion -- by embedding the rotation group $SU(2)$ for $j=3/2$ spin within a higher dimension group, such as $SO(4)$ in $d=4$.

Alternatively, Abrikosov \cite{Abrikosov:1974qv} constructed a series of spin-orbit coupled models models with quadratic band touching in arbitrary dimensions $d$ using Clifford matrices. The Hamiltonian is

\begin{equation}
H =  \gamma^{a} d_{a}(\bk), \qquad \qquad a = 1, 2, \ldots, (d-1)(d+2)/2
\end{equation}

\noindent where $\gamma^{a}$ are anti-commuting Clifford matrices. The $d_{a}(\bk)$ are functions determined by requiring that the dispersion should be quadratic [$H^{2} =  d_{a}(\bk)d_{a}(\bk) = c^{2}k^{4}$]  and by imposing an orthogonality condition over angular integration [$\int \ud\Omega_\bk d_{a}(\bk)d_{a}(\bk') = 0$ for $\bk'$ constant]. Since one can construct the $SO(N)$ Clifford algebra from sets of Pauli matrices, it is easy to determine the dimension of the representation $r_d$ -- in particular, $r_3 = 4$, $r_{4} = 16$. From the $\gamma^{a}$, we define generalized $A_{ij}$ matrixes satisfying

\begin{equation}\label{eq:AmatrixAbr}
\{A_{ij},A_{kl}\} = \frac{d}{d-1}(\delta_{ik}\delta_{jl} + \delta_{il}\delta_{jk}) - \frac{2}{d-1}\delta_{ij}\delta_{kl}
\end{equation}

\noindent This relationship is also well defined in fractional dimensions and is useful for the $\vare$-expansion since only this combination appears in the RG calculation. In certain dimensions, a $\gamma$ matrix will not enter the Hamiltonian and thus give a notion of chirality to the states. Importantly, this does not occur for the relevant dimensions for our model $d=3,4$. This does occur $d=2$, where the Hamiltonian describes, e.g.~a special case of the quadratic chiral edge states at the surface of a crystalline topological insulator \cite{Fu:2011fk}.  In $d=3$, the relationship of states around the $\Gamma_{8}$ point to the $SO(5)$ Clifford algebra has been discussed in detail by \cite{Murakami:2004qf}.

\subsubsection{Renormalization Group Fixed Point}

Here we briefly give the results of the RG flow in the $\vare$-expansion. The action is 

\begin{equation}
S = \int\!\!\ud^{3} x \left[\psi^{\dag} i \partial_{t} \psi - c (\partial_{i}\psi^{\dag})A_{ij}(\partial_{j}\psi) + \frac{1}{8\pi e^{2}} (\partial_{i}\varphi) (\partial_{i}\varphi) - \varphi \psi^{\dag}\psi \right]
\end{equation}

\noindent where we have decoupled the Coulomb interaction into an auxiliary field $\varphi$. This is the scalar part of the full electromagnetic interaction, up to a rescaling of the field by the coupling constant $e$. Taking the scaling dimensions to be $[K] = - [L] = 1$, the tree level scaling is $[\omega] = - [T] = 2, [e^{2}] = 4-d = \vare$. We perform the RG in such a way that the kinetic term $ck^{2}$ remains unchanged during scaling. As a result one obtains the RG equation for $e^{2} / c$, which has a stable fixed point. Using the structure \eqref{eq:AmatrixAbr} the fixed point value is

\begin{equation}
\alpha^{*} = \left(\frac{e^2}{c}\right)^* = \frac{24\pi}{3r_{d}+2}\vare \to \frac{12 \pi}{25} 
\end{equation}

\noindent The anomalous scaling exponent is

\begin{equation}
z = 2 - \frac{2}{3r_{d}+2}\vare \to 2 - \frac{1}{25}
\end{equation}

\noindent where we set $\vare = 1$ and $r_{4} = 16$, which is the smallest dimension of the representation in $d=4$.

If we were to have chosen to formally use the $d=3$ structure \eqref{eq:AmatrixApp} of the commutation relation but still perform the integrals in $d = 4 - \vare$ the results would be:

\begin{equation}
\alpha^{*} = \left(\frac{e^2}{c}\right)^* = \frac{64 \pi}{3(3r_d + 2)} \vare \to \frac{32 \pi}{21}, \qquad z = 2 - \frac{9}{4(3r_d +2)}\vare \to 2 - \frac{9}{56}
\end{equation}

\noindent with $r_{3} = 4$. The difference in the above results is mostly due to the different dimension of the representation $r_{3,4}$, rather the form of the commutator $\{A_{ij},A_{kl}\} $ or the prescription of  angular integration. For calculating the shear viscosity, we shall therefore use the $d=3$ structure \eqref{eq:AmatrixApp} with $r_{3} = 4$; all the results in the main text are evaluated with this choice.

\subsection{Kinetic Equation with Collisions}

\subsubsection{Eigenstates in $d=3$}

For the later discussion of the kinetic equation approach, it will be convenient to temporarily work in  $d=3$, with the physical spin structure. The eigenstates of the free Hamiltonian $H_{0} = A_{ij} k_{i} k_{j}$ are plane waves with spin

\begin{equation}
[\psi_{m}(\bx)]_{a} = \frac{e^{i\bk\cdot\bx}}{\sqrt{V}} \cdot  \mathscr{D}^{3/2}_{a m}(\varphi,\theta,0) 
\end{equation}

\noindent where $\mathscr{D}^{j}_{m'm}(\varphi,\theta,\chi)$ are Wigner D-symbols; $(\varphi,\theta)$ are the angular coordinates of $\bk$ and the choice $\chi = 0$ of our eigenstates is just a phase convention. Here $a$ labels the band index which we will usually suppress. The index $m$ labels both the species of particles ($\vert m \vert = 3/2$ conduction electrons, $\vert m \vert = 1/2$ valance holes) and their helicity ($\sigma = \sgn m$). Since the projection of the angular momentum onto the $\bk$ is different for particles and holes, we can see that the system does not posses true particle-hole symmetry. We can write our fields $\psi$ as

\begin{equation}
\psi(\bx) = \int\!\!\frac{\ud^{3} k}{(2\pi)^{3}} \left[{u}_{\sigma}(\bk) {c}_{\sigma}(\bk) e^{i \bk\cdot \bx} + {v}_{\sigma}(\bk) {h}_{\sigma}^{\dag}(\bk) e^{-i \bk\cdot \bx} \right]
\end{equation}
\noindent where $[{u}_{\pm}]_{a}(\bk)  = \mathscr{D}^{3/2}_{a,\pm3/2}(\varphi,\theta,0), [{v}_{\pm}]_{a}(-\bk)  =  \mathscr{D}^{3/2}_{a,\pm1/2}(\varphi,\theta,0)$. 

\noindent The Hamiltonian is

\begin{equation}
H = \int\!\!\frac{\ud^{3} k}{(2\pi)^{3}} \epsilon_{k}  \left[ c^{\dag}_{\sigma}(\bk)c_{\sigma}(\bk)+h^{\dag}_{\sigma}(\bk)h_{\sigma}(\bk)\right], \qquad \epsilon_{k} = ck^{2},
\end{equation}

\noindent the charge current is ${j}_{i} = {j}_{i}^{(1)}+{j}_{i}^{(2)}$, where

\begin{eqnarray}
{j}_{i}^{(1)} & = & 2 e c \int\!\!\frac{\ud^{3}k}{(2\pi)^{3}} \ {k}_{i}  \left[ c^{\dag}_{\sigma}(\bk)c_{\sigma}(\bk)-h^{\dag}_{\sigma}(\bk)h_{\sigma}(\bk)\right] \\
{j}_{i}^{(2)} & = &  2 e c \sum_{\sigma = \pm}\sqrt{\frac{3}{2}}\int\!\!\frac{\ud^{3}k}{(2\pi)^{3}} \ k  \left( \sigma T_{\sigma}(\hat{\vect{e}}_{i})
c_{\sigma 3/2}^{\dag}(\bk)c_{\sigma 1/2}(\bk) +h.c. \right)
\end{eqnarray}

\noindent and the integrated stress tensor contribution of the electrons and holes is ${T}_{ij} = {T}^{(1)}_{ij}  + {T}^{(2)}_{ij}$, where

\begin{eqnarray}
\label{eq:tij1} {T}^{(1)}_{ij} &=& c \int\!\!\frac{\ud^{3}k}{(2\pi)^{3}}\  2k_{i}k_{j} \left[ c^{\dag}_{\sigma}(\bk)c_{\sigma}(\bk)+h^{\dag}_{\sigma}(\bk)h_{\sigma}(\bk)\right] \\
\label{eq:tij2} {T}^{(2)}_{ij} &=& c \sqrt{\frac{3}{2}} \sum_{\sigma = \pm} \int\!\!\frac{\ud^{3}k}{(2\pi)^{3}} \ k \left( \left[\sigma  T_{\sigma}(\hat{\vect{e}}_{i}) k_{j} + \sigma T_{\sigma}(\hat{\vect{e}}_{j}) k_{i}\right] c_{\sigma 3/2}^{\dag}(\bk)c_{\sigma 1/2}(\bk) + h.c. \right).
\end{eqnarray}

\noindent In the above, $\hat{\vect{e}}_{i}$ is the unit vector in the $i$ direction and we have introduced the matrix element

\begin{equation}
T_{+}(\bhn) = T_{-}^{\dag}(\bhn) =  \frac{[\hat{\vect{e}}_{-} \cdot \bhk ]  (-i \bhk\times\bhn -  \bhn + (\bhk\cdot\bhn)\bhk  ) \cdot \hat{\vect{z}} }{1 - (\bhk\cdot\hat{\vect{z}})^{2}}, \qquad \hat{\vect{e}}_{-} = \tfrac{1}{\sqrt{2}} (\hat{\vect{x}}-i\hat{\vect{y}})
\end{equation}

\subsubsection{General Set-Up}

Within the approximations discussed in the main text, the kinetic equation with interaction reduces to the semi-classical kinetic equation for the diagonal distribution functions

\begin{equation}
\left( \partial_{t} + {\vect{v}}_{\bk} \cdot \partial_{\bx} + {\vect{F}}_{ext} \cdot \partial_{\bk} \right) f^{a}(t,\bx,\bk) = - C[f^{a}]
\end{equation}

\noindent The collision term is 

\begin{multline}
C[f] =  \frac{1}{2} \sum_{bcd} \int_{\bp',\bk,\bk'}  \vert M^{cd}_{ab}(\bp,\bk,\bp',\bk') \vert^{2}  (2\pi)^{d+1} \delta(\vare_{\bp} + \vare_{\bk} - \vare_{\bp'} -\vare_{\bk'}) \delta^{d}(\bp + \bk -\bp' - \bk') \\ 
\left\{ f^{a}({\bp}) f^{b}({\bk}) [ 1-f^{c}({\bp'})][1-f^{d}({\bk'})] - f^{c}({\bp'}) f^{d}({\bk'})   [ 1-f^{a}({\bp})][1-f^{b}({\bk})]\right\}   
\end{multline}

\noindent The matrix terms are given by the tree-level scattering processes $ab \to cd$. As stated in the main text, we shall neglect scattering processes in the s-channel (involving particle-hole annihilation and reformation). To find the transport coefficients, we look for perturbations about equilibrium 

\begin{equation}\label{eq:fulldistfunc}
f^{a}(\bk) = f_{0}(\bk) + \beta [1-f_{0}(\bk)] f_{0}(\bk) \chi^{a}(k) I_{ij\dots}(\bk) X_{ij\dots}
\end{equation}

\noindent where $X_{ij\ldots}$ is the external driving force appropriate for the transport coefficient under consideration, $I_{ij\dots}(\bk)$ is tensor appropriate for the symmetry of the driving force and $\chi(k)$ the perturbed distribution function. We also introduce the short-hand $\chi_{ij\ldots}(\bk) = I_{ij\ldots}(\bk) \chi(k)$. We shall keep this generality for the moment, and focus on the case of shear viscosity later.  For both particles and holes, the equilibrium distribution function is $f_{0}(\bk) = {1}/[{\exp(c k^{2} / T) +1}]$. 
Explicitly

\begin{multline}
Q[\chi] =  - \sum_{a} \int_{\bk} f_{0}(\bk) [1-f_{0}(\bk)] \beta q^{a} \chi^{a}(k) 
-\frac{\beta}{16} \sum_{abcd} \int_{\bk,\bp,\bk',\bp'} \vert M^{cd}_{ab}(\bp,\bk,\bp',\bk') \vert^{2}  f_{0}(\bp) f_{0}(\bk) [ 1-f_{0}(\bp')][1-f_{0}(\bk')]  \\
(2\pi)^{d+1} \delta(\vare_{\bp} + \vare_{\bk} - \vare_{\bp'} -\vare_{\bk'}) \delta^{d}(\bp + \bk -\bp' - \bk') \cdot 
  \left\{ \chi^{a}_{ij\dots}(\bp) + \chi^{b}_{ij\dots}(\bk) - \chi^{c}_{ij\dots}(\bp') - \chi^{d}_{ij\dots}(\bk')\right\}^{2}
\end{multline}

\noindent  where we used detailed balance to make manifest that $C$ is hermitian. Here $q^{a}$ are `charge factors'; in the case of the shear viscosity  $q^{a} = k^{4}$ for all species. The above rewrite also gives rise to the factor $1/8$ to stop overcounting of  $a \leftrightarrow b$ and $ab \leftrightarrow cd$ processes. 

We maximize $Q[\chi]$ by projecting the function $\chi(k) = \sum_{\mu} a_{\mu} F_{\mu}(k)$ onto a set of basis functions $F_{\mu}(k)$. Maximizing $Q$ then reduces to the solution of matrix equation 

\begin{equation}\sum_{\nu} C_{\mu \nu} a_{\nu} = b_{\mu}\end{equation}

 \noindent for $a_{\nu}$.  From now on  we  set $d=4$, since we are working in the $\vare$-expansion and interesting in the dominant $\vare$ term. We also denote by  $\bP,\bK,\bQ$ the scaled momenta $\bP = \bp \sqrt{c/T}$. We defin $\bq$ such that  $\bp' = \bp+\bq$ and $\bk' = \bk -\bq$; we perform the $\bk'$ integral to remove the delta function in momentum and replace the $\bp'$ integration by a $\bq$ integration. We rewrite the delta function enforcing energy conservation during scattering as
 
\begin{eqnarray}
\nonumber \delta(\vare_{\bp} + \vare_{\bk} - \vare_{\bp'} -\vare_{\bk'}) &=& \frac{2Q}{c} \cdot \frac{c}{T} \int_{-\infty}^{\infty}\!\!\ud y \ \delta(K^{2} - \vert \bK - \bQ \vert^{2} - 2Qy) \delta (P^{2}- \vert \bP + \bQ\vert^{2} + 2Q y) \\
&= & \frac{2 Q}{T \cdot (2KQ) (2PQ)} \int_{y_{0}}^{y_{1}}\!\!\ud y \ \delta\left(\cos \theta_{KQ} - \frac{2y +Q  }{2K}\right) \delta\left(\cos \theta_{PQ} - \frac{2y-Q }{2P}\right)
\end{eqnarray}

\noindent where $y_{0} = \min[P + Q/2, K - Q/2]$ and $ y_{1} = \max[-P + Q/2, -K - Q/2]$ are restrictions imposed so that the cosines are defined. 

Unlike usual `leading-q' expansions \cite{Arnold:2000vl}, we only modify the matrix elements -- dropping the spin structure as well as particle-hole creation processes -- but do not expand the functions $\chi_{ij}(\bp'),\chi_{ij}(\bk')$ in $\bq$ nor set $\bq=0$ in the Fermi functions. After performing the angular integrals analytically, we are left with a four-dimensional integral (over $k,q,p,y$) to calculate $C_{\mu\nu}$, which we perform numerically using the cubature algorithm  Cuhre of the Cuba library \cite{Hahn:2005fk}.

\subsubsection{Shear Viscosity}

To find the shear viscosity we apply a divergence-free background flow pattern of momentum $\bP(\bx)$ with $\partial_{k} P_{k} = 0$:

\begin{eqnarray}
X_{ij} &=& \frac{1}{2} \left( \Pdd{x_{i}}{P_{j}} + \Pdd{x_{j}}{P_{i}}\right)  \\
I_{ij}(\bk) &=&  \sqrt{\frac{d}{d-1}}\left({k}_{i}{k}_{j} - \frac{1}{d}\delta_{ij}k^{2}\right)
\end{eqnarray}

\noindent and solve the kinetic equation for $\chi$. The shear viscosity can then be found from the stress tensor

\begin{equation}
T_{ij}= - \eta\left(\delta_{ik}\delta_{jl}+\delta_{il}\delta_{jk}\right) 2 c X_{ij} =\sum_{a} \int_{\bk}2c k_{i} k_{j} f^{a}(\bk)
\end{equation}

\noindent Due to the symmetry of the diagonal part of the stress tensor under particle and hole exchange, the function $\chi(k)$ will be the same for particles and holes. The charge factors $q^{a} = k^{4}$ for all species. In the low-q approximation for a matrix element, the sum entering the kinetic equation functional is $\sum_{abcd}  \vert M^{cd}_{ab}(\bp,\bk,\bp',\bk') \vert^{2} = r_{d}^{2} \cdot (4\pi \alpha^{*} c / q^{2})^{2}$, taking into account the  exchange symmetry and over-counting factors. We define the dimensionless scaling function

\begin{equation}
\Phi(K) = \frac{T(\alpha^{*})^{2}}{c} \cdot  \chi\left(k\sqrt{\frac{T}{c}}\right)
\end{equation}

\noindent which we expand onto a set of basis functions $\Phi(K) = \sum_{\mu} a_{\mu} F_{\mu}(K)$. We choose

\begin{equation}\label{eq:varbasis}
F_{\mu}(K) = \left[1+e^{-K^2}\right]^{ 2}L^{(3)}_{\mu}(K^{2})
\end{equation}

\noindent where $L^{(3)}_{\mu}(x)$ are associated Laguerre polynomials. The choice of Laugerre polynomials is well known from the case of transport in a Boltzmann gas \cite{Lifshitz:1981fk} -- it simplifies the form of $b_{\mu}$ to

\begin{equation}
b_{\mu} = \frac{r_{4}}{16 \pi^{2} (\alpha^{*})^{2}  } \left(\frac{T}{c}\right)^{2} \frac{18}{\sqrt{3}} \delta_{\mu,0}
\end{equation}

\noindent due to the orthogonality properties of the Laguerre polynomials. These basis functions also give a reasonably well conditioned matrix $C_{\mu\nu}$  which can easily be evaluated and inverted numerically. We use a set of $12$ basis functions.  Performing the variational solution gives

\begin{equation}
\eta = - \frac{r_{d}}{16 \pi^{2} (\alpha^{*})^{2} } \left(\frac{T}{c}\right)^{2} \frac{1}{3\sqrt{3}} \int_{0}^{\infty}\!\!\ud K \ f_{0}(K)[1-f_{0}(K)]  K^{7} \Phi(K) = -\frac{r_{d}}{16 \pi^{2} (\alpha^{*})^{2}  } \left(\frac{T}{c}\right)^{2} \frac{a_{0}}{\sqrt{3}} \simeq  \frac{3.14 }{(\alpha^{*})^{2}} \left(\frac{T}{c}\right)^{2}
\end{equation}

\noindent with $r_{d} = 4$. This is the result given in the main text.

\subsection{Coulomb Contribution to Stress Tensor}

In addition to the contribution of the electrons and holes given in \eqref{eq:tij1} and \eqref{eq:tij2}, the stress tensor has a contribution from the Coulomb interaction

\begin{equation}\label{eq:StressIntTensor}
T^{(int)}_{ij} = - \frac{1}{4\pi e^2}\int \!\!\frac{\ud^{4}k}{(2\pi)^{4}} \ \left[ k_{i} k_{j} - \frac{1}{2}\delta_{ij} k^{2} \right] \langle \varphi(\bk) \varphi(-\bk)\rangle
\end{equation}

\noindent where by the average $\langle \varphi(k) \varphi(-k)\rangle$ we mean the semi-classical limit of the correlator evaluated in the non-equilibrium case. We will argue in our approach that the above term is of higher order in $\vare$ and can be neglected in our calculation of $\eta$. Since the Coulomb interaction is non-dynamic, the contribution to the viscosity when we apply a background momentum $\bP(\bx)$ will be through loop corrections involving the fermions. We shall use the results we found from the Boltzmann equation so that our answer is self-consistent. We can show in Keldysh perturbation theory \cite{Lifshitz:1981fk, Altland:2010uq}, that for the scalar interaction

\begin{equation}
\langle\varphi(x_{1})\varphi(x_{2}) \rangle = - D^{R}(x_{1},x_{3}) \Pi^{-+}(x_{3},x_{4}) D^{A}(x_{4},x_{2})
\end{equation}

\noindent where $D^{R,A}$ are the retarted and advanced Green functions of the scalar field and $\Pi^{-+}$ the appropriate self-energy. Here we use the notation of \cite{Lifshitz:1981fk} for the contour labels. We now perform a semi-classical expansion, retaining only the lowest order derivatives of $\bP$

\begin{equation}\label{eq:semiclassical}
\langle\varphi(\bk)\varphi(-\bk) \rangle = - D^{R}(\bk) \Pi^{-+}(\bx,\bk) D^{A}(\bk) + \frac{i}{2}\left[ \Pdd{k_{i}}{D^{R}(\bk)} \Pdd{x_{i}}{\Pi^{-+}(\bx,\bk)} D^{A}(\bk) - {D^{R}(\bk)} \Pdd{x_{i}}{\Pi^{-+}(\bx,\bk)} \Pdd{k_{i}}{D^{A}(\bk) } \right] + \ldots
\end{equation}

\noindent where further terms in the semi-classical expansion are higher derivative of $\bP$ and can be neglected. At our level of approximation we include the thermal screening term and the lowest order $k^{2}$ term with the fixed point interaction

\begin{equation}
D^{R,A}(k) = \frac{1}{\nu r_{d} T c^{-2} + k^{2} / (4\pi c \alpha^{*})}
\end{equation}

\noindent where $\nu$ is a numerical constant. Since the electron contribution to the viscosity goes as $\sim 1 / \vare^{2}$ we need the dominant power of $\langle\varphi(k)\varphi(-k) \rangle$ in $\vare$ to be no higher than $\sim 1/\vare$ in order for \eqref{eq:StressIntTensor} to be important. The second term in \eqref{eq:semiclassical} does not contribute -- to see this consider that the spatial derivative on $\Pi^{-+}$, which acts only on $\bP(\bx)$ in the equilibrium distribution function of the fermions. To order $\pdd{x_{i}}{P_{j}}$, $\pdd{x_{i}}{\Pi^{-+}}$ does not contribute any powers of $\vare$, while the $k$ derivatives of $D^{R,A}$ only contribute positive powers of $\vare$.

\noindent In our approximation of considering only the diagonal distribution functions discussed in the main text, the semi-classical self-energy in the first term of \eqref{eq:semiclassical} is 

\begin{equation} 
\Pi^{-+}(\bx,\bk) = - \sum_{a}\int \!\!\frac{\ud^{4}q}{(2\pi)^{4}} f_{a}(\bq) [ 1- f_{a}(\bk+\bq)] \delta(\vare_{\bk+\bq} -\vare_{\bq})
\end{equation}

\noindent where $f_{a}$ are the non-equilibrium distribution functions \eqref{eq:fulldistfunc}. To order $\pdd{x_{i}}{P_{j}}$, the $f$ contribute a single $\sim 1/\vare^{2}$, which are, however, compensated. For a generic momentum $k \sim \sqrt{T}$ in \eqref{eq:StressIntTensor}, we may neglect the screening in $D^{R,A}$ and each propagator contributes a power $\vare$. For a soft momentum on the order of the screening scale or above $k \sim \sqrt{\vare T}$, the contribution to \eqref{eq:StressIntTensor} is suppressed by powers of $k$ from $\sim k^{2} d^{4}k$ in the integral. To lowest order in $\vare$, we can therefore neglect \eqref{eq:StressIntTensor} in our calculation of $\eta$.

\end{document}